# Symmetry breaking at the (111) interfaces of SrTiO$_3$ hosting a 2D-electron system

G. M. De Luca[1,2], R. Di Capua[1,2], E. Di Gennaro[1,2], A. Sambri[1,2], F. Miletto Granozio[2], G. Ghiringhelli[3,4], D. Betto[5], C. Piamonteze[6], N.B. Brookes[5] and M. Salluzzo[2]

[1]*Dipartimento di Fisica, Università "Federico II" di Napoli, Complesso Monte Sant'Angelo via Cinthia, I-80126 Napoli, Italy*

[2]*CNR-SPIN, Complesso Monte Sant'Angelo via Cinthia, I-80126 Napoli, Italy*

[3]*Dipartimento di Fisica, Politecnico di Milano, Piazza Leonardo da Vinci 32, I-20133 Milano, Italy*

[4]*CNR-SPIN, Politecnico di Milano, Piazza Leonardo da Vinci 32, I-20133 Milano, Italy*

[5]*ESRF, 71 Avenue des Martyrs, 38000 Grenoble, France*

[6]*Swiss Light Source, Paul Scherrer Institut, CH-5232 Villigen PSI, Switzerland*

We used x-ray absorption spectroscopy to study the orbital symmetry and the energy band splitting of (111) LaAlO$_3$/SrTiO$_3$ and LaAlO$_3$/EuTiO$_3$/SrTiO$_3$ heterostructures, hosting a quasi two-dimensional electron system (q2DES), and of a Ti-terminated (111) SrTiO$_3$ single crystal, also known to form a q2DES at its surface. We demonstrate that the bulk tetragonal Ti-3d $D_{4h}$ crystal field is turned into trigonal $D_{3d}$ crystal field in all cases. The symmetry adapted a$_{1g}$ and e$^\pi_g$ orbitals are non-degenerate in energy and their splitting, Δ, is positive at the bare STO surface but negative in the heterostructures, where the a$_{1g}$ orbital is lowest in energy. These results demonstrate that the interfacial symmetry breaking induced by epitaxial engineering of oxide interfaces has a dramatic effect on their electronic properties, and it can be used to manipulate the ground state of the q2DES.

*Introduction* Transition metal oxides are renowed for the novel physical phenomena emerging at their surfaces and interfaces, like the formation of a quasi-two-dimensional electron system (q2DES) at the interface between (001)[1], (110) and (111)[2] SrTiO$_3$ (STO) single crystals and LaAlO$_3$ thin films (LAO). Some of the distinctive properties of the (001) LAO/STO q2DES arise from the occurrence of an orbital reconstruction, firstly demonstrated by x-ray linear dichroism (XLD)[3] and later on by Angle Resolved Photoemission Spectroscopy (ARPES).[4,5,6,7] In the (001) LAO/STO, this orbital reconstruction causes a reverse ordering and splitting of the bulk conduction bands derived from the non-degenerate t$_{2g}$ (3d$_{xy}$, 3d$_{xz}$, 3d$_{yz}$) orbitals of Ti 3d-states (Fig. 1a). A similar orbital reconstruction was observed also in other (001) titanate heterostructures, like in the (001) LaAlO$_3$/EuTiO$_3$/SrTiO$_3$ (LAO/ETO/STO) system, which hosts a spin-polarized q2DES.[8,9] In the (110) LAO/STO, instead, the q2DES does not show the reverse ordering of the bands as it happens in the the (001) cases[10,11]. However, in both systems, the bulk tetragonal crystal field, D$_{4h}$, associated to the characteristic distorted octahedral oxygen-cages around Ti-ions, is retained also at the interface. Thus, the overall q2DES's electronic properties can still be described within a D$_{4h}$ frameworks.

The D$_{4h}$ symmetry is on the other hand naturally broken at the (111) SrTiO$_3$ surface, where the crystal field becomes trigonal owing to the D$_{3d}$ symmetry (Fig. 1b) associated to the hexagonal surface lattice[12]. Within the D$_{3d}$ symmetry, the t$_{2g}$ states are mixed, so that the three usual orbitals transform into a$_{1g}$ and e$^\pi_g$ states (Fig. 1c-d), which are futher split by the crystal field and/or by the spin-orbit coupling[13]. Consequently, the band structure of (111) oxide heterostructures is susceptible of being more significantly altered at their interfaces respect the (001) and (110) cases. As matter of fact, interesting phenomena were theoretically predicted in (111) oxides, like non-trivial topological effects and the possibility of quantum Hall effect up to room temperature[14]. In particular, a massive symmetry breaking was predicted in the case of (111) LAO/STO quantum wells, with ordering of the ($a_{1g}$, $e^{\pi}_g$) derived bands depending on the strain and on the confinement[12].

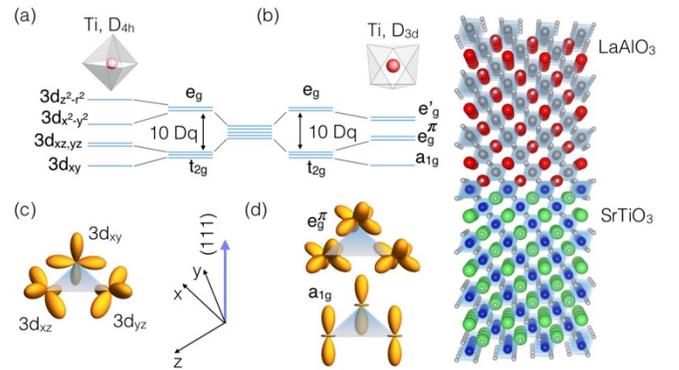

Fig. 1: Energy diagram of the 3d Ti states in the case of (a) tetragonal and (b) trigonal crystal fields. (c) t$_{2g}$ *3d$_{xy}$, 3d$_{xz}$, 3d$_{yz}$* orbitals and (d) *a$_{1g}$* and *e$^\pi_g$* orbitals viewed from the (111) surface (shaded triangles). On the right a schematic of the (111) LAO/STO heterostructure.

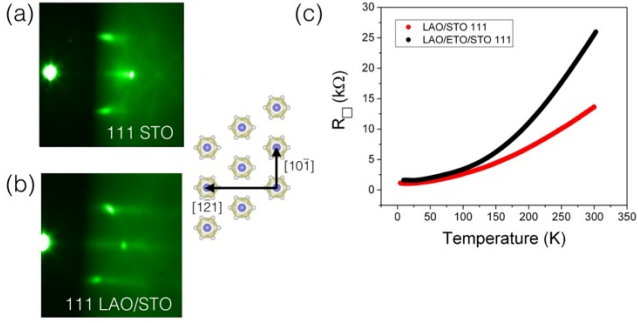

Fig. 2: RHEED image along the [$10\bar{1}$] crystallographic direction for (a) (111) STO and (b) (111) LAO/STO samples. (c) Sheet resistance of the (111) LAO/STO (red) and (111) LAO/ETO/STO (black) as function of the temperature.

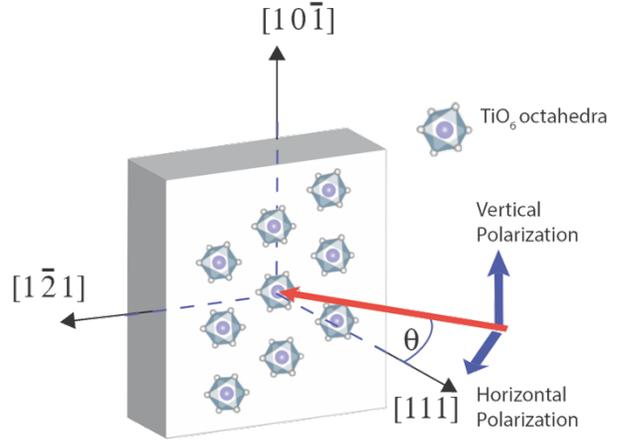

Fig. 3: Schematic of the XAS and XLD experimental configuration. The x-rays propagation vector (red arrow) forms an angle θ with the [111] surface normal. Blue lines indicate the direction of the corresponding polarization vectors. The [$10\bar{1}$] crystalline direction vector is parallel to the vertical polarization for any incidence angle. On the surface plane we have superimposed the schematic arrangement of the Ti (violes spheres) and oxygen (gray spheres).

ARPES studies of the (111) STO surface state showed a Fermi surface interpreted as originating from the projection on the (111) plane of the confined and degenerate $3d_{xz}$, $3d_{yz}$, and $3d_{xy}$ bands [15, 16]. Since the surface unit cell of (111) STO has an inherent trigonal symmetry, the derived surface states have to be described as a combination of $t_{2g}$ orbitals, thus in terms of $a_{1g}$ or $e^{\pi}_g$ derived bands. One would then expect a band structure very different from the bulk. Consequently, it is natural to ask ourselves if the orbital and the electronic symmetry of the Ti 3d-bands in (111) heterostructures and at the (111) STO surface are perturbed by surface/interface symmetry breaking, and how this would affect the global properties of the q2DES formed thereby.

Here we have determined the orbital symmetry and splitting of the Ti-3d states at the (111) STO surface and at the (111) LAO/STO and LAO/ETO/STO interfaces by measuring, in total electron yield (TEY), the titanium $L_{2,3}$-edge and the oxygen K-edge x-ray absorption spectra (XAS) and related x-ray linear dichroism (XLD). The TEY XAS and XLD spectra probe the interface electronic properties within a depth of about 3 nm from the interface. In all the (111) heterostructures we find that a $D_{3d}$ symmetry replaces the tetragonal $D_{4h}$ symmetry of bulk SrTiO3. Moreover, the $a_{1g}$ and $e^{\pi}_g$ orbitals are non-degenerate, and their splitting depends sensitively on the specific system analyzed, so that it is opposite in sign in the heterostructures compared to the bare (111) STO surface state.

*Exeperimental Results* The XAS measurements were performed at the ID32 beamline of the European Synchrotron Radiation Facility (ESRF) [17] and at the X-treme beamline of the Swiss Light Source (SLS) [18]. In both cases, the samples were measured in a variable temperature cryostat in ultra-high vacuum. Ti-terminated (111) STO single crystals were prepared by buffered hydrofluoric acid chemical etching and oxygen annealing, and were characterized by a 1x1 surface lattice (Fig. 2a). The LAO/STO and LAO/ETO/STO heterostructures were grown by pulsed laser deposition assisted by reflection high energy electron diffraction[19, 20] onto the (111) STO single crystals mentioned above (Fig. 2b), using a background oxygen pressure of $1 \times 10^{-4}$ mbar and laser fluence of 1.5 J/cm$^2$. In particular, we studied heterostructures characterized by 10 unit cells (uc) of LAO in the case of (111) LAO/STO, and by 3 uc of ETO plus 10 uc of LAO sequentially deposited on STO in the case of (111) LAO/ETO/STO.[20] In both types of samples, we observed a metallic temperature dependence of the sheet resistance down to 4.2 K, as shown in Fig. 2c [20]. The low temperature values of the sheet resistances were comparable to standard (001) LAO/STO and LAO/ETO/STO heterostructures deposited in the same deposition conditions.

The samples were mounted on the XAS setups with the (111) surface in the vertical laboratory plane, and the vertical axis (V) parallel to the [$10\bar{1}$] crystallographic direction determined by x-ray diffraction, i.e. parallel to one of the three equivalent Ti-Ti bonds within the hexagonal surface lattice (Fig 3). The energy resolution at both beamlines was around 100 meV. X-ray linear dichroism spectra were obtained by averaging at least 16 XAS spectra for each of the two linear horizontal (H) and vertical (V) polarizations. Linear dichroism spectra were measured as function of the angle θ between the incident photons and the surface normal (Fig. 3). While for the vertical polarization the x-ray electric field is always parallel to the surface and to the $[10\bar{1}]$ crystallographic direction (Fig. 3), with horizontal polarization, the electric field vector is in the (111) surface plane and along the $[1\bar{2}\bar{1}]$ direction for θ=0 (Normal incidence, NI), and almost parallel to the [111] lattice vector, i.e. along the surface normal, for θ =70° (Grazing incidence, GI).

Table 1: Crystal field parameters used to reproduce the XAS spectra of the heterostructures studied.

| 10Dq (eV) | slater (%) | HWHM Peak A (eV) | HWHM Peak B (eV) | HWHM Peak C (eV) | HWHM Peak D (eV) |
|---|---|---|---|---|---|
| 2.23 | 80 | 0.1 | 0.34 | 0.5 | 0.55 |

The XAS and XLD experimental spectra were compared to theoretical spectra obtained by atomic multiplet scattering calculations using the CTM4XAS code.[21]

In Fig. 4a we show a comparison between the Ti- $L_{2,3}$ edge XAS spectra of the three systems. We can see that the spectra on the three different systems analyzed are all very similar, apart a very slight increase (less than 5%) of the half width at half maximum (HWHM) Lorentian broadenings of the main peaks in the case of the (111) LAO/STO and LAO/ETO/STO heterostructures as compared to the (111) STO single crystal. The XAS spectra can be reproduced assuming a $3d^0$ atomic configuration ($Ti^{4+}$) using the parameters listed in Table 1, namely the value of 10Dq, i.e. the overall splitting between $e_g$ and $t_{2g}$ orbitals, the Slater integrals, and the Lorentzian broadenings (HWHM) of the four peaks in the XAS spectra (Fig. 3a). In Fig. 3a we used a $D_{3d}$ symmetry, however similar XAS spectra are obtained using a $D_{4h}$ symmetry.

In Fig. 4b we show the angle dependence of the $XLD = I_H - I_V$ dichroic spectra measured at 5 K of the LAO/ETO/STO q2DES. We can see that while the XLD intensity changes as function of the angle, its shape remains substantially constant from 35° to 70°. This result suggests that the out-of-plane dichroism, which is mainly probed in gracing incidence, dominates the $I_H - I_V$ spectral shape. In normal incidence, θ =0°, on the other hand, $I_H - I_V$ is much smaller and has a different shape, since it is related to in plane linear dichroism $XLD_\parallel = I_{[1\bar{2}\bar{1}]} - I_{[10\bar{1}]}$.

From the spectra acquired between 35° and 70°, we can determine the experimental out-of-plane dichroism,

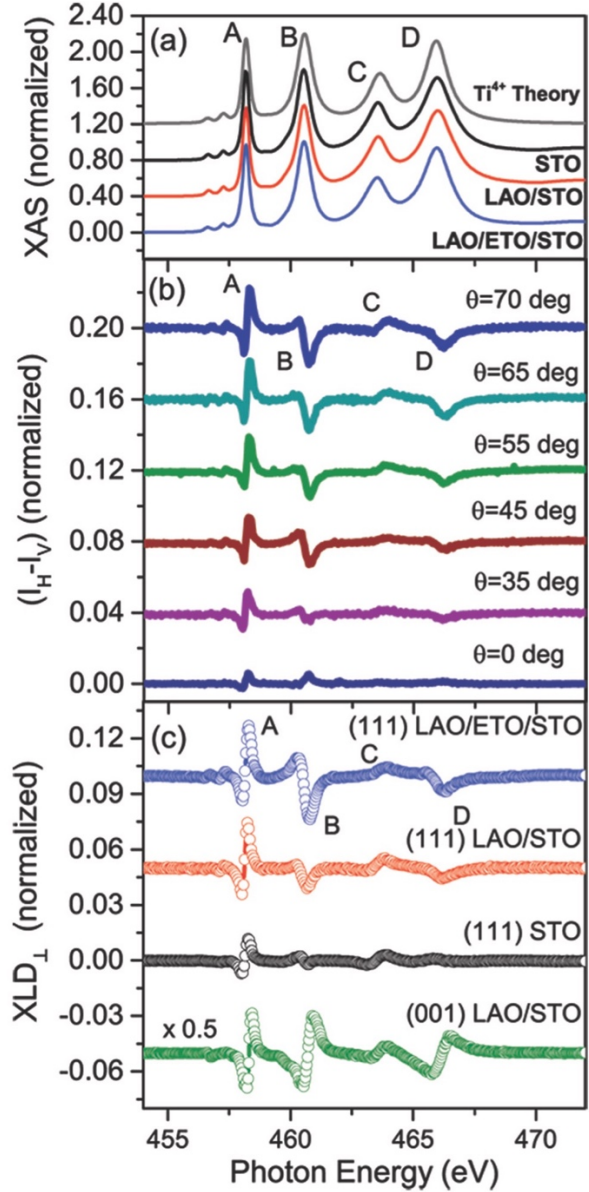

Fig. 4: Total electron Yield XAS spectra of (111) STO (green line) LAO/STO (red line) and LAO/ETO/STO (black line) measured at 5 K in grazing incidence conditions. The dark gray line (upper curve in the figure) assuming a pure $Ti^{4+}$ valence and using the parameters listed in Table 1. (b) $I_H$-$I_V$ data in the same conditions as function of the angle θ measured on the (111) LAO/ETO/STO sample. Features A, B, C and D indicate the main absorption peaks at $L_3$ (A, B) and $L_2$ (C, D). (c) $XLD_\perp$ data, normalized to the XAS maxima at $L_3$ edge of (001) LAO/STO (green cicles, divided by a factor 2), (111) STO (black circles), (111) LAO/STO (red circes) and (111) LAO/ETO/STO (blue circles). The data are shifted by a constant factor (0.05) for clarity.

$XLD_\perp = I_{[111]} - I_{[10\bar{1}]}$, considering that $I_{[10\bar{1}]} = I_V$ and $I_{[111]} \cong (I_H - I_V cos^2\vartheta)/sin^2\vartheta$ [22]. As shown in the suppl. mat.,[20] the $XLD_\perp$ data obtained from different angles coincide, confirming that the dichroic signal is due to the in-plane/out-of plane anisotropy. Moreover, the $XLD_\perp$ spectral shape of (111) heterostructures, shown in Fig. 4c, differs significantly from that measured previously on (001) and (110) LAO/STO, and on bulk (001) STO.[3,10] In particular, the three selected cases of (111) STO heterostructures and the (001) LAO/STO $XLD_\perp$ differ both in the amplitude and in the shape of the main features (A, B, C, D).

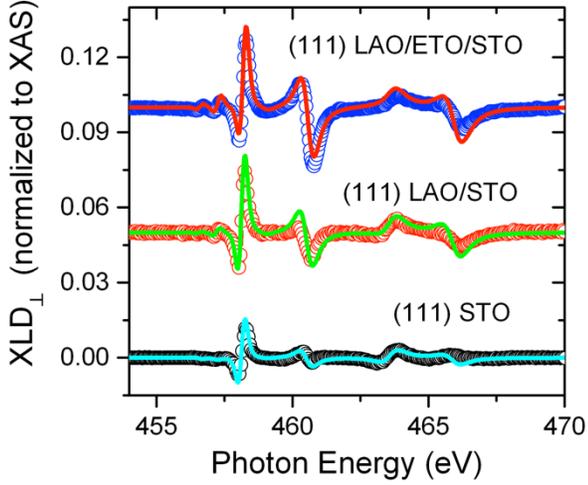

FIG. 5: Comparison between out of plane linear dichroism spectra measured in total electron yield of (111) STO (black circles), LAO/STO (red circles) and LAO/ETO/STO (blue circles). Continuous lines are the best fit of the data by atomic multiplet calculations

Table 2: $D_\sigma$ and $D_\tau$ parameters obtained from atomic multiplet calculations reproducing the XLD data of Fig. 4 and the integral of the XLD normalized to the integral of the sum $I_{[111]} + I_{[10\bar{1}]}$.

| Sample | $D_\sigma$ (meV) | $D_\tau$ (meV) | $\Delta$ (meV) | XLD (x10$^{-3}$) |
|---|---|---|---|---|
| STO | 2±0.3 | -2.2±0.1 | 8±3 | 0.1±0.1 |
| LAO/STO | 12±1 | -4.2±0.2 | -8±3 | -3.0±0.5 |
| LAO/ETO/STO | 16±1 | -4.2±0.2 | -20±3 | -6.0±0.5 |

Before comparing the data to theoretical calculations, we discuss what would be the expected $XLD_\perp$ in case of $D_{4h}$ symmetry, i.e, in a case where the bulk symmetry of STO is preserved up to the interface/surface unit cells probed by TEY-XLD (around 3 nm). Then the $XLD_\perp$ would be a combination of the $I_{[001]}$ and $I_{[100]}$ Ti $L_{2,3}$ XAS along the [001] and [100] directions, calculated in the $D_{4h}$ symmetry. For example, the out of plane dichroism, $XLD_{\perp,111,D4h}$, would be given by [20]:

$$XLD_{\perp,111,D4h} = I_{[111]} - I_{[10\bar{1}]} = \frac{1}{6}\left(I_{[100]} - I_{[001]}\right). \quad (1)$$

while the in plane $XLD_{\parallel,111,D4h}$ dichroism would be:

$$XLD_{\parallel,111,D4h} = I_{[1\bar{2}1]} - I_{[10\bar{1}]} = \frac{1}{3}\left(I_{[100]} - I_{[001]}\right). \quad (2)$$

Thus, if the tetragonal symmetry is preserved, with consequent degeneracy of the $t_{2g}$ states, the dichroism would be zero also for the (111) STO surface. Moreover, even in the case of non-degenerate $t_{2g}$ states, the non-vanishing $XLDs$ would have opposite sign with respect to the $XLD_\perp = I_{[001]} - I_{[100]}$ of (001) heterostructures, and their shape would be similar to the one measured in the case of (110) LAO/STO[10]. The data show clearly that this is not case, thus suggesting a breaking of symmetry for all the three systems analyzed. In particular, the features B and D in the spectra have a shape reversed compared to the the (001) LAO/STO interface, while the features A and C are similar in shape. Furthermore, the overall amplitude of the dichroism is smaller in (111) heterostructures.

In order to understand the effective symmetry of the novel electronic configuration, we fit the XAS and the $XLD_\perp$ spectra using atomic multiplet scattering calculations by the CTM4XAS code. Starting from the XAS analysis shown in Figure 4a, we then calculate the $XLD_\perp$ spectra of 3d$^0$ Ti$^{4+}$ ions, in the case of a $D_{3d}$ symmetry for various values of the $a_{1g}$ - $e^\pi_g$ splitting, depending on the crystal field terms $D_\tau$ and $D_\sigma$, the only parameters optimized in the fitting. More in details, it is possible to show that the energy splitting $\Delta = E(a_{1g}) - E(e^\pi_g)$ is given by [20, 23]:

$$\Delta = \sqrt{\Gamma_{e_g} - 5D_q - 5/2\, D_\sigma - 7/2\, D_\tau}. \quad (3)$$

with

$$\Gamma_{e_g} = (5D_q)^2 + (3/2\, D_\sigma)^2 + (5/2\, D_\tau)^2 + \\ -5D_q D_\sigma + 25/3\, D_q D_\tau - 15/2\, D_\sigma D_\tau. \quad (4)$$

It turns out that peaks A and C are roughly influenced by $D_\tau$, while B and D depend mainly on $D_\sigma$. For (111) STO the best agreement is for $D_\tau$ = -2.2 meV and $D_\sigma$ = 2 meV, corresponding to $\Delta$ = +8 meV, implying that the orbitals of $e^\pi_g$ symmetry are the lowest in energy. We remind that the $a_{1g}$ orbital is an equal weight linear combination of the pristine $3d_{xz}$ $3d_{yz}$ and $3d_{xy}$ orbitals with real coefficients, while for the $e^\pi_g$ the coefficients are complex numbers. Therefore, the orbital symmetry is quite different from the one expected in the case of bulk $t_{2g}$ states simply projected on the (111) surface. The opposite case, is the one of the (111) LAO/ETO/STO (and LAO/STO) heterostructure, where we find $\Delta$ = -20 (-8) meV, i.e. the lowest energy lying orbitals have $a_{1g}$ symmetry.

It is important to underline that from the analysis it emerges that STO-based heterostructures are characterized by an electronic configuration which is close to a perfect $t_{2g}^0$ ($3d^0$) occupation. However, the integral of the $XLD_\perp$ spectra is not null, and in particular we find it small and slightly positive in the case of the (111) STO surface, while it is more than one order of magnitude larger, and negative in sign, in the case of the heterostructures, as shown in Table 2. The negative $XLD_\perp$ integral suggests that the (few) 3d electrons in the (111) q2DES preferentially occupy bands derived by orbitals that have a predominant out-of-plane character, i.e. $a_{1g}$ derived bands in both (111) LAO/STO and LAO/ETO/STO, in agreement with the overall $XLD_\perp$ data.

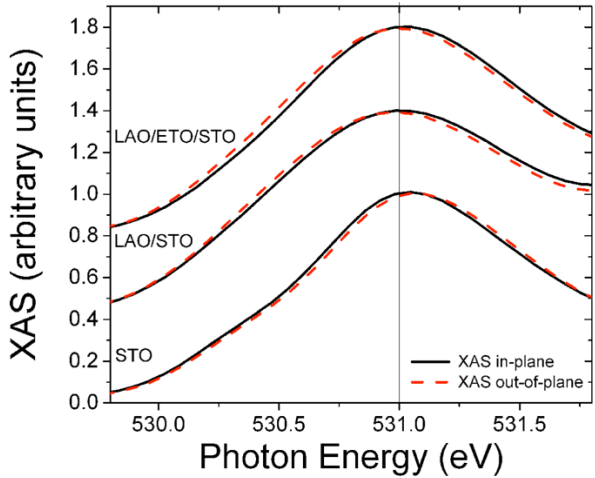

FIG. 6. Comparison between in plane (black lines) and out of plane (dashed red line) oxygen K-edge XAS spectra measured in total electron yield at 4.5 K of (111) STO, LAO/STO and LAO/ETO/STO samples. The data on different samples are displaced by a constant factor 0.4 and are normalized to the XAS peak at 531 eV.

The removal of the 3d degeneracy and the consequent anisotropy in the electronic states is also reflected in the relative contributions of the in-plane and out-of-plane orbitals to the conduction band of the surfaces/interfaces. This behavior was determined from oxygen K-edge XAS spectra measured with the interface sensitive TEY method as function of the linear polarization of the photons (Fig. 6). In this x-ray absorption process, 1s core electrons belonging to the oxygen ions are excited to the 2p states. Since the oxygen 2p states are nominally filled in both LAO, ETO and STO, a 1s→2p transition occurs only because of the hybridization of the 2p states with the d-orbitals of the cations in the system. Because of this hybridization, the peak at 531 eV of Fig. 6 is related to the conduction band, located in STO just above the Fermi level and it is composed, in these (111) heterostructures, of 3d titanium $a_{1g}$ and $e_g^\pi$ orbitals hybridized with oxygen 2p states. By using the TEY method of detection and grazing incidence conditions, the only contribution to XAS spectra in the 529÷533 eV region comes from the TiO2 interface layer[24]. In the (111) STO case, we find that the oxygen K-edge XAS, measured with linear polarization parallel to the surface, is slightly at lower energy respect the one measured with perpendicular polarization. In the heterostructures, on the other hand, the energy shift between the in-plane and out of plane XAS is reversed and, in agreement with the Ti-$L_{2,3}$ edge spectra, it is maximum in the (111) LAO/ETO/STO heterostructures.

*Discussion and conclusions*
According to theory[12], LAO/STO quantum wells can host a variety of unconventional ground states, ranging from spin, orbital polarized (with selective $e_g^0$, $a_{1g}$, or $d_{xy}$ occupation), Dirac point Fermi surfaces, to charge-ordered at band phases, as function of the strain and confinement. The realizaition of these phases, however, depends on the effective breaking of the $D_{4h}$ symmetry into a $D_{3d}$ symmetry, and on the effective magnitude of the trigonal crystal field, which is required to open a gap between otherwise degenerate $t_{2g}$ manifolds. In this work, we demonstrated that the q2DES formed in (111)-STO based heterostructures has indeed a $D_{3d}$ orbital symmetry, as opposite to the bulk $D_{4h}$ symmetry, and that the trigonal crystal field is suffciently strong to open a gap between $a_{1g}$ and $e_g^\pi$ states.

These results have, therefore, interesting consequences about the current understanding of the physics of (111) STO based q2DES. Transport measurements have shown that the (111) LAO/STO q2DES condenses to a superconducting state below a critical temperature, Tc, of 250 mK[25], and analogously to the (001) LAO/STO[26], it is characterized by a quite sizable spin-orbit coupling[27], with a dome-shape Tc vs carrier density. Furthermore, recently an accurate analysis of the gate voltage dependence of the Hall effect has given evidences of a transition from one-band to two-bands electron transport in the same system[28]. The interpretation of all these results is not easily reconciled with a band structure derived by the simple projection of degenerate $D_{4h}$ $t_{2g}$ bands on the (111) plane, which would result in an effective one-electron band transport scenario. Our spectroscopic study, on the other hand, suggests that a transition from a one- to a two-bands regime is plausible because the (111) LAO/STO q2DES is characterized by at least two non-degenerate electron-like bands. It is worth noting that recent transport data proposed an out of plane spin-texture of the q2DES[29] in electrostatically gated (111) STO, consistent with a breaking of the bulk $D_{4h}$ crystal field into a trigonal crystal field (C3 symmetry), in overall agreement with our findings.

Morevoer, our data show that some of the crucial ingredients needed to realize quantum Hall effect systems and non-trivial topological states, as proposed theoretically for (111) perovskite oxide heterostructures[14], could be actually present in the case of (111) STO-based q2DES. In particular, we demonstrated that the trigonal crystal field in (111)-STO based heterostructures is sufficiently strong to open a gap up to 20 meV between $a_{1g}$ and $e_g^\pi$ states. We also showed that this gap can be changed, and even inverted through epitaxial engineering. Since the LAO/STO q2DES is characterized by a sizable Rashba spin-orbit coupling in both (001)[26] and (111)[27] cases, this study opens interesting perspectives for the exploration of novel unconventional ground states in (111) STO based heterostructures.

**ACKNOWLEDGMENTS**

The Authors acknowledge received funding from the project Quantox of QuantERA ERA-NET Cofund in Quantum Technologies (Grant Agreement N. 731473) implemented within the European Union's Horizon 2020 Programme".

G. M. De Luca[1,2], R. Di Capua[1,2], E. Di Gennaro[1,2], A. Sambri[1,2], F. Miletto Granozio[2], G. Ghiringhelli[3,4], D. Betto[5], C. Piamonteze[6], N.B. Brookes[5] and M. Salluzzo[2]

[1] *Dipartimento di Fisica, Università "Federico II" di Napoli, Complesso Monte Sant'Angelo via Cinthia, I-80126 Napoli, Italy*

[2] *CNR-SPIN, Complesso Monte Sant'Angelo via Cinthia, I-80126 Napoli, Italy*

[3] *Dipartimento di Fisica, Politecnico di Milano, Piazza Leonardo da Vinci 32, I-20133 Milano, Italy*

[4] *CNR-SPIN, Politecnico di Milano, Piazza Leonardo da Vinci 32, I-20133 Milano, Italy*

[5] *ESRF, 71 Avenue des Martyrs, 38000 Grenoble, France*

[6] *Swiss Light Source, Paul Scherrer Institut, CH-5232 Villigen PSI, Switzerland*

*(Dated: June 27, 2018)*

## Supplementary Note 1

In the D$_{4h}$ symmetry, in the simple ortho-normal base *[100]*, *[010]*, *[001]*, the angle $\alpha_{hk}$ between two vectors **h** and **k** characterized by indices *[h$_x$, h$_y$, h$_z$]* and *[k$_x$, k$_y$, k$_z$]* respectively, satisfies the relation (from $\mathbf{h}\cdot\mathbf{k} = |\mathbf{h}||\mathbf{k}|\cos\alpha_{hk}$):

$$\cos^2\alpha_{hk} = \frac{(h_x k_x + h_y k_y + h_z k_z)^2}{(h_x^2 + h_y^2 + h_z^2)(k_x^2 + k_y^2 + k_z^2)} \quad (S1)$$

Therefore, the XAS intensity along a generic direction $\mathbf{n} = (n_x, n_y, n_z)$ can be written as:

$$I_{nx,\,ny,\,nz} = I_{[1,0,0]}\cos^2\alpha_x + I_{[0,1,0]}\cos^2\alpha_y + I_{[0,0,1]}\cos^2\alpha_z \quad (S2)$$

where $\alpha_x, \alpha_y, \alpha_z$, are the angles formed by $\mathbf{n}$ and the base vectors, and:

$$\cos^2\alpha_j = \frac{n_j^2}{n_x^2 + n_y^2 + n_z^2} \quad (S3)$$

If the bulk symmetry of STO were preserved up to the interface/surface, it should happen that $I_{[1,0,0]} = I_{[0,1,0]} \neq I_{[0,0,1]}$, and therefore we could write:

$$XLD_{\perp,111,D4h} = I_{[111]} - I_{[10\bar{1}]} = \frac{1}{3}(I_{[100]} + I_{[010]} + I_{[001]}) - \frac{1}{2}(I_{[100]} + I_{[001]}) =$$
$$= -\frac{1}{6}I_{[100]} + \frac{1}{3}I_{[010]} - \frac{1}{6}I_{[001]} = \frac{1}{6}(I_{[100]} - I_{[001]})$$
(S4)

$$XLD_{\parallel,111,D4h} = I_{[1\bar{2}1]} - I_{[10\bar{1}]} = \frac{1}{6}I_{[100]} + \frac{2}{3}I_{[010]} + \frac{1}{6}I_{[001]} - \frac{1}{2}(I_{[100]} + I_{[001]}) =$$
$$= -\frac{1}{3}I_{[100]} + \frac{2}{3}I_{[010]} - \frac{1}{3}I_{[001]} = \frac{1}{3}(I_{[100]} - I_{[001]})$$
(S5)

In our experimental setup, the direction of Linear Vertical (LV) polarization of the incident light is always parallel to the $[10\bar{1}]$ crystallographic direction, independently on the angle $\theta$ between the normal to the sample surface (i.e., the $[111]$ crystallographic direction) and the propagation vector of the incident light. On the contrary, the direction of Linear Horizontal (LH) polarization of the incident light forms an angle $\theta$ with the $[1\bar{2}1]$ crystallographic direction, lying in the $[1\bar{2}1]$-$[111]$ plane (cfr. Fig. S1).

As a consequence:

$$\begin{cases} I_V = I_{[10\bar{1}]} & \forall \theta \\ I_H = I_{[1\bar{2}1]}\cos^2\theta + I_{[111]}\sin^2\theta \end{cases} \Rightarrow I_V - I_H = I_{[10\bar{1}]} - \left(I_{[1\bar{2}1]}\cos^2\theta + I_{[111]}\sin^2\theta\right)$$ (S6)

Being $XLD_\perp = I_{[111]} - I_{[10\bar{1}]}$ and $XLD_\parallel = I_{[1\bar{2}1]} - I_{[10\bar{1}]}$, from relations (S6) it easily comes out that:

$$I_H - I_V = XLD_\parallel \cos^2\theta + XLD_\perp \sin^2\theta$$ (S7)

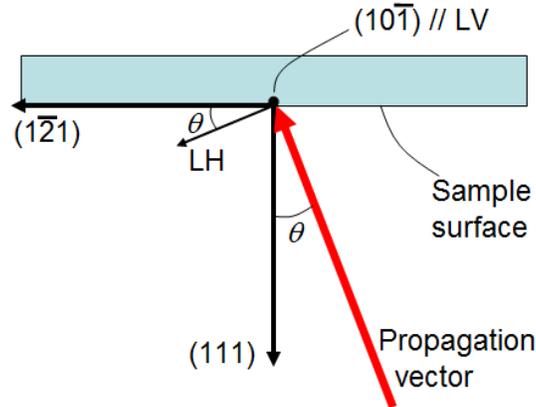

**Fig. S1**: a sketch illustrating the relation between the LH polarization and the crystallographic direction as a function of the angle q between the propagation vector of light and the normal to the sample surface.

In Normal Incidence ($\theta = 0$), (S7) gives $I_H - I_V = XLD_\parallel$, as expected. On the other hand, for $\theta = 90°$, from (S7) we would have $I_H - I_V = XLD_\perp$. At $\theta = 70°$ (Grazing Incidence condition in our experiment) we can compute from (S7): $I_H - I_V = 0.883\ XLD_\perp + 0.117\ XLD_\parallel$, which means, as stated in the main text, that at Grazing Incidence we mainly probe the out-of-plane dichroism.

In addition, from (S7) we can infer:

$$XLD_\perp = \frac{I_H - I_V}{\sin^2 \theta} - XLD_\parallel \tan^2 \theta \qquad (S8)$$

which can be approximated at $(I_H - I_V)/\sin^2\theta$, being in-plane XLD much smaller than out-of-plane XLD.

In Fig. S2 we show $XLD_\perp$ for LAO/ETO/STO (111) heterostructure determined from the measured $I_H$-$I_V$. It is possible to note that the result is very similar for each value of θ from 35° to 70°.

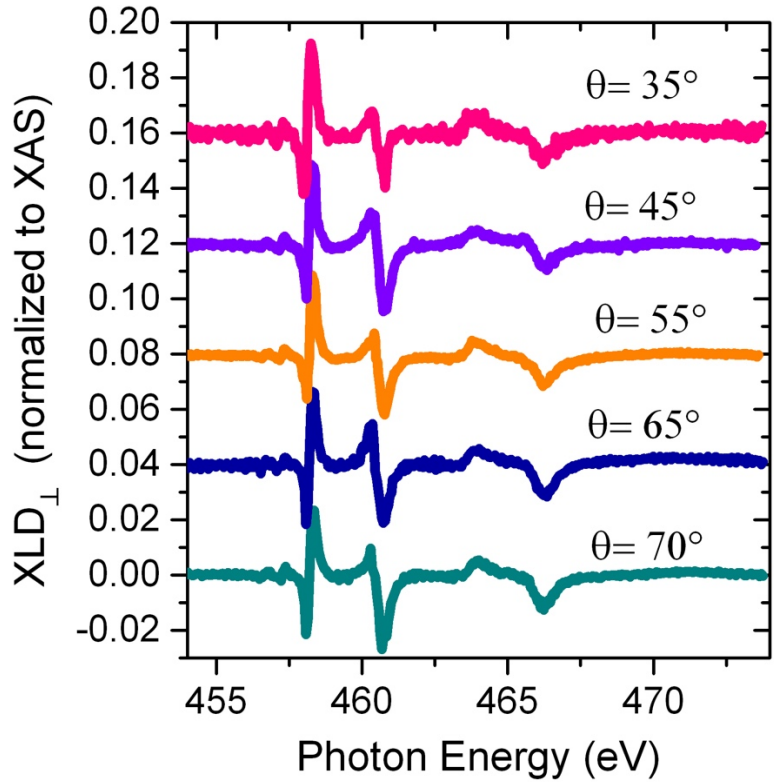

**Fig. S2**: Out of plane dichoism normalized to the XAS intensity at $L_3$, of the (111) LAO/ETO/STO heterostructure determined from the measured dichroism at different angles θ from 35° (top) to 70° (bottom). The data are displaced by a constant factor 0.04.

# Supplementary note 2: Calculation of the XAS and XLD spectra using a trigonal crystal field

If we quantize the orbitals along the (111) direction, *i.e.* the z axis is directed along (111), we have the following orbitals:

$$a_{1g} = d_0$$
$$e_g^{\pi\pm} = \sqrt{\frac{2}{3}}d_{\pm 2} \mp \sqrt{\frac{1}{3}}d_{\mp 1} \tag{S9}$$
$$e_g'^{\pm} = \sqrt{\frac{1}{3}}d_{\pm 2} \pm \sqrt{\frac{2}{3}}d_{\mp 1}$$

where $d_{ml}$ are the spherical harmonics $Y_2^{ml}$ quantized along the (111) direction.

Using this convention, we have, for trigonal crystal field $V_t$:

$$\langle d_{\pm 2}|V_t|d_{\pm 2}\rangle = -\frac{2}{3}Dq + 2D_\sigma - D_\tau$$
$$\langle d_{\pm 1}|V_t|d_{\pm 1}\rangle = -\frac{8}{3}Dq - D_\sigma + 4D_\tau$$
$$\langle d_0|V_t|d_0\rangle = -4Dq - 2D_\sigma - 6D_\tau \tag{S10}$$
$$\langle d_{\pm 2}|V_t|d_{\mp 1}\rangle = \langle d_{\mp 1}|V_t|d_{\pm 2}\rangle = \pm\frac{10}{3}\sqrt{(2)}Dq$$

Therefore, we have:

$$\langle a_{1g}|V_t|a_{1g}\rangle = -4Dq - 2D_\sigma - 6D_\tau$$
$$\langle e_g^{\pi\pm}|V_t|e_g^{\pi\pm}\rangle = -4Dq + D_\sigma + \frac{2}{3}D_\tau$$
$$\langle e_g'^{\pm}|V_t|e_g'^{\pm}\rangle = 6Dq + \frac{7}{3}D_\tau \tag{S11}$$
$$\langle e_g^{\pi\pm}|V_t|e_g'^{\pm}\rangle = \sqrt{2}D_\sigma - \frac{5\sqrt{2}}{3}D_\tau$$

The orbitals are not eigenfunctions of the system anymore. As you can see, there is a non-diagonal matrix element that mixes the two types of $e_g$ orbitals (obviously). The $a_{1g}$ stands alone and it is an eigenfunction.

To be accurate one should diagonalise the Hamiltonian for the $e_g$ states and find eigenfunctions and eigenvalues.

Hamiltonian:

$$\begin{array}{c} \\ e_g^\pi \\ e_g' \end{array} \begin{array}{cc} e_g^\pi & e_g' \end{array} \left[\begin{array}{cc} -4Dq + D_\sigma + \frac{2}{3}D_\tau & \sqrt{2}Dq - \frac{5\sqrt{2}}{3}D_\tau \\ \sqrt{2}Dq - \frac{5\sqrt{2}}{3}D_\tau & 6Dq + \frac{7}{3}D_\tau \end{array}\right] \tag{S12}$$

Eigenvalues:

$$E_1 = \frac{1}{6}\left(6Dq + 3D_\sigma + 9D_\tau - \sqrt{3}\sqrt{300Dq^2 - 60DqD_\sigma + 100DqD_\tau + 27D_\sigma^2 - 90D_\sigma D_\tau + 75D_\tau^2}\right)$$
$$= Dq + \frac{1}{2}D_\sigma + \frac{3}{2}D_\tau - \sqrt{\Gamma_{e_g}} \tag{S13}$$
$$E_2 = \frac{1}{6}\left(6Dq + 3D_\sigma + 9D_\tau + \sqrt{3}\sqrt{300Dq^2 - 60DqD_\sigma + 100DqD_\tau + 27D_\sigma^2 - 90D_\sigma D_\tau + 75D_\tau^2}\right)$$
$$= Dq + \frac{1}{2}D_\sigma + \frac{3}{2}D_\tau + \sqrt{\Gamma_{e_g}}$$

with

$$\Gamma_{e_g} = 25Dq^2 + \frac{9}{4}D_\sigma^2 + \frac{25}{4}D_\tau^2 - 5DqD_\sigma + \frac{25}{3}DqD_\tau - \frac{15}{2}DqD_\tau \qquad (S14)$$

Eigenvectors in base $e_g^\pi, e_g'$:

$$\Phi_1 = \left\{ \frac{-\sqrt{3}\sqrt{300Dq^2 - 60DqD_\sigma + 100DqD_\tau + 27D_\sigma^2 - 90D_\sigma D_\tau + 75D_\tau^2} - 30Dq + 3D_\sigma - 5D_\tau}{2\sqrt{2}(3D_\sigma - 5D_\tau)}, 1 \right\}$$

$$= \left\{ \frac{-6\sqrt{\Gamma_{e_g}} - 30Dq + 3D_\sigma - 5D_\tau}{2\sqrt{2}(3D_\sigma - 5D_\tau)}, 1 \right\}$$

$$\Phi_2 = \left\{ \frac{+\sqrt{3}\sqrt{300Dq^2 - 60DqD_\sigma + 100DqD_\tau + 27D_\sigma^2 - 90D_\sigma D_\tau + 75D_\tau^2} - 30Dq + 3D_\sigma - 5D_\tau}{2\sqrt{2}(3D_\sigma - 5D_\tau)}, 1 \right\}$$

$$= \left\{ \frac{+6\sqrt{\Gamma_{e_g}} - 30Dq + 3D_\sigma - 5D_\tau}{2\sqrt{2}(3D_\sigma - 5D_\tau)}, 1 \right\} \qquad (S15)$$